\documentclass[11pt]{article}
\usepackage{moriond,epsfig,pifont,floatflt}

\def\Journal#1#2#3#4{{#1} {\bf #2}, #3 (#4)}

\def\NPB{{\em Nucl.\ Phys.\ }B}
\def\PLB{{\em Phys.\ Lett.\ }B}

\def\ZPC{{\em Z.\ Phys.\ }C}
\def\JHEP{{\em JHEP}}
\def\PRC{\em Phys.\ Reports\/}

\def\be{\begin{equation}}
\def\ee{\end{equation}}
\def\bea{\begin{eqnarray}}
\def\eea{\end{eqnarray}}
\def\PT{\mbox{\scriptsize{PT}}}
\def\NP{\mbox{\scriptsize{NP}}}
\def\lrang#1{\left\langle#1\right\rangle}
\def\cM{{\cal{M}}}
\def\cA{{\cal A}}

\newcommand{\epem}{\rm e^+\rm e^-}
\def\as{\alpha_s}
\def\cO#1{{\cal{O}}\left(#1\right)}
\def\d{\hat{d}}

\def\wisd#1#2#3{\begin{flushright}
\begin{minipage}{#1in}
\begin{itemize}
\item[\ding{'53}] 
#2
\end{itemize}
\vspace{-0.6cm}
\begin{flushright}
#3
\end{flushright}
\end{minipage} 
\end{flushright}
}

\begin{document}
\vspace*{4cm}
\title{Perturbative QCD and Power Corrections \\[4mm]
{\normalsize Invited talk at the International Conference \\
``Frontiers of  Matter'', Blois, France, June 1999}}

\author{ Yu.L. Dokshitzer  }

\address{INFN sezione di Milano, via G. Celoria 16, 20133 Milan, Italy\\
St.Petersburg Nuclear Physics Institute, Gatchina, 18350, Russia}

\maketitle\abstracts{A short review is given of the present status of
  the studies of genuine confinement effects in multiple
  hadron production in hard processes.}

\section{Beware of Soft Confinement}
\wisd{3.5}{
What I tell you three times is true.} 
{``{\em The Hunting of the Snark}''. 
Lewis Carroll}
\vspace{3mm}

Perturbative QCD (pQCD) covers orders of magnitude in the basic hard
cross sections. 
Scattered clouds on the pQCD horizon (presently, a ``high-$p_T$
anomaly'' and some problems with hadroproduction of heavy quarks and
direct photons) don't define the weather.  However, viewing the QCD
landscape it is essential to remember that its ``peace and quite'' is
deceptive.  The closest analogy which comes to mind comparing QCD
with its electroweak SM-counterpart is that of hell and heaven: the
former being scary but entertaining, the latter --- perfect but boring
(for a detailed review see~\cite{Dante}).

\subsection{The name of the game: MLLA}
\wisd{3.7}
{Although this may seem a paradox, all exact science is
  dominated by the idea of {approximation}.}
 {Bertrand Russell}
 \vspace{3mm}

 In spite of the smallness of the coupling at small distances,
 $\as\ll1$, quarks and gluons willingly multiply in hard interactions.
 This happens because the actual parameter of the \PT\ expansion gets
 enhanced by the $\log$ of the scale of the large momentum $Q$ applied
 to the system: $ \as(Q) \> \mbox{\ding{'334}}\> \as(Q)\log Q \,
 \sim\,1 $.  Such ${\log}$-enhanced contributions have to be taken
 care of in all orders, giving rise to ``leading-log resummations''.
 The structure of these contributions allows for a {\em probabilistic
   parton interpretation}.  Parton multiplication in jets is described
 by the so-called Modified Leading Logarithmic Approximation (MLLA)
 which embodies the exact angular ordering resulting from coherence in
 multiple soft gluon radiation. As an approximation, MLLA is necessary
 for deriving asymptotically correct \PT\ predictions.

\newpage
 
 QCD coherence is crucial for treating particle multiplication {\bf
   inside} jets, as well as for hadron flows {\bf in-between} jets.
 Moreover, it allows the prediction of the {\em shape}\/ of the
 inclusive energy distributions practically from the ``first
 principles'', apart from an overall unknown normalization {\bf
   constant}.  \hfill{\large\ding{'54}}

\subsection{Inclusive energy spectra}

It is well known that the DIS structure functions cannot be calculated
perturbatively. What pQCD controls is the scaling violation pattern,
governed by the QCD parton evolution equation which describes
how the parton densities change with changing the scale of the 
transverse-momentum probe:
\be 
\frac{\partial}{\partial \ln k_\perp} D(x,k_\perp) = \frac{\as(k_\perp)}{\pi}
\, P(z) \otimes D(x/z,k_\perp) \>.
\ee 
In the Mellin moment space, 
$f_\omega\equiv\int_0^1 f(x)x^{\omega-1}dx$, 
the equation becomes algebraic, yielding
\be\label{yd_sfandim} 
D_\omega(k_\perp) \equiv
D_\omega(Q_0)\cdot
\exp\left\{ \int^{k_\perp}_{Q_0} \frac{dk}{k} \gamma_\omega(\as(k))\right\},
\quad \gamma_\omega(\as) = \frac{\as}{\pi} \, P_\omega \>.
\ee
It is the $\omega$-dependence of the input function 
$D_\omega(Q_0)$ (``initial parton distributions'') that limits predictability 
of the Bjorken-$x$ dependence of DIS cross sections.

In the time-like jet evolution, due to the Angular Ordering,
the evolution equation becomes non-local in the $k_\perp$ space:
\be 
\frac{\partial}{\partial \ln k_\perp} D(x,k_\perp) = \frac{\as}{\pi}
\, P(z) \otimes D(x/z,\, z\cdot k_\perp)\,; \qquad
\left[\,\mbox{\bf AO:}\>\> \Theta'=k_\perp'/zE\,\ge\, 
\Theta=k_\perp/E \right] .
\ee 
Using the Taylor expansion trick,
\be
  D(x/z,\, z\cdot k_\perp) =  \exp\left\{ \ln z \frac{\partial}{\partial \ln
      k_\perp} \right\}  D(x/z,k_\perp)\>=\>
z^{\frac{\partial}{\partial \ln  k_\perp}} \cdot D\>,
\ee
the solution in the moment space comes out similar to that for the DIS
case, Eq.~\ref{yd_sfandim}, 
but now with an {\em operator}\/ as an argument of the splitting
function $P$:
\be 
\d\cdot D_\omega = \frac{\as}{\pi} \, P_{\omega+\d} \cdot D_\omega\>,
\quad \d\equiv \frac{\partial}{\partial \ln k_\perp}\>.  
\ee 
This leads to the differential equation 
\be\label{yd_difeq} 
\left(  P^{-1}_{\omega+\d}\>\d \>-\> \frac{\as}{\pi}
  -\left[P^{-1}_{\omega+\d}, \frac{\as}{\pi}\right]P_{\omega+\d}
\right) \cdot D =0\>.  
\ee 
Since we are interested in the small-$x$
region, the essential moments are small, $\omega\ll1$. For the sake of
illustration let us keep only the most singular piece of the splitting
function (DLA), 
\be\label{yd_andimapp} 
P_\omega \simeq \frac{2N_c}{\omega}\>.  
\ee 
Then Eq.~\ref{yd_difeq} immediately gives
a quadratic equation for the anomalous dimension, 
\be\label{yd_dlaeq}
(\omega+\gamma)\gamma \>-\> \frac{2N_c\as}{\pi}
+\cO{\frac{\as^2}{\omega}} \>=\>0\,.  
\ee 
NB: It suffices to use the
next-to-leading approximation to the splitting function,
$P_\omega \simeq {2N_c}/{\omega} -a$, with 
$a = {11N_c}/{6}+ {n_f}/(3N_c^2)$,  
and to keep the leading correction coming from differentiation 
of the running coupling in Eqs.~\ref{yd_difeq},~\ref{yd_dlaeq}, 
to get the more accurate MLLA anomalous dimension $\gamma_\omega$.

The leading anomalous dimension following from Eq.~\ref{yd_dlaeq} is
\be\label{yd_andimdla}
 \gamma = {\scriptsize \frac\omega 2}
 \left(-1+ \sqrt{1+ {8N_c\as}/{\omega^2}}\right). 
\ee
When expanded to the first order in $\as$, it coincides with that for
the space-like evolution, $\gamma\simeq\as/\pi\cdot P_\omega$, with 
$P$ given in Eq.~\ref{yd_andimapp}.

The time-like DLA anomalous dimension Eq.~\ref{yd_andimdla} (as well as
its MLLA improved version) has a curious property. Namely, in a sharp
contrast with the DIS case, it allows the momentum integral in
Eq.~\ref{yd_sfandim} to be extended to very small scales.  Even
integrating down to $Q_0=\Lambda$, the position of the ``Landau pole''
in the coupling, one gets a finite answer for the distribution (the
so-called {\em limiting spectrum}), simply because the $\sqrt{\as(k)}$
singularity happens to be integrable!

It would have been a bad taste to actually trust this formal
integrability, since the very perturbative approach to the problem
(selection of dominant contributions, parton evolution picture, etc)
relied on $\as$ being a numerically small parameter. 
However, the important thing is that, due to time-like coherence
effects, the (still perturbative but ``smallish'') scales, 
where $\as(k)\gg\omega^2$, contribute to $\gamma$ basically in a
$\omega$-{\em independent}\/ way, 
$ 
 \gamma+{\omega}/{2} \propto \sqrt{\as(k)} \neq f(\omega)
$.
This means that ``smallish'' momentum scales $k$ affect only an
overall {\em normalization}\/ without affecting the {\em shape}\/ of the
$x$-distribution.
Since such is the r\^ole of the ``smallish'' scales, it is natural 
to expect the same for the truly small --- non-perturbative ---
scales where the partons transform into the final hadrons. 
This idea has been formulated as a hypothesis of local parton-hadron
duality (LPHD).\cite{LPHD,KO} 

According to LPHD, the $x$-shape of the so-called ``limiting''
spectrum which one obtains by formally setting $Q_0=\Lambda$ in the
parton evolution equations, should be mathematically similar to that
of the inclusive hadron distribution.  Another essential property is
that the ``conversion coefficient'' should be a true constant
independent of the hardness of the process producing the jet under
consideration.  Starting from the LEP-I epoch, this ``prediction''
stood up to scrutiny by $\epem$, DIS and Tevatron experiments.

\vfill

\noindent
\begin{minipage}{0.5\textwidth}
\psfig{figure=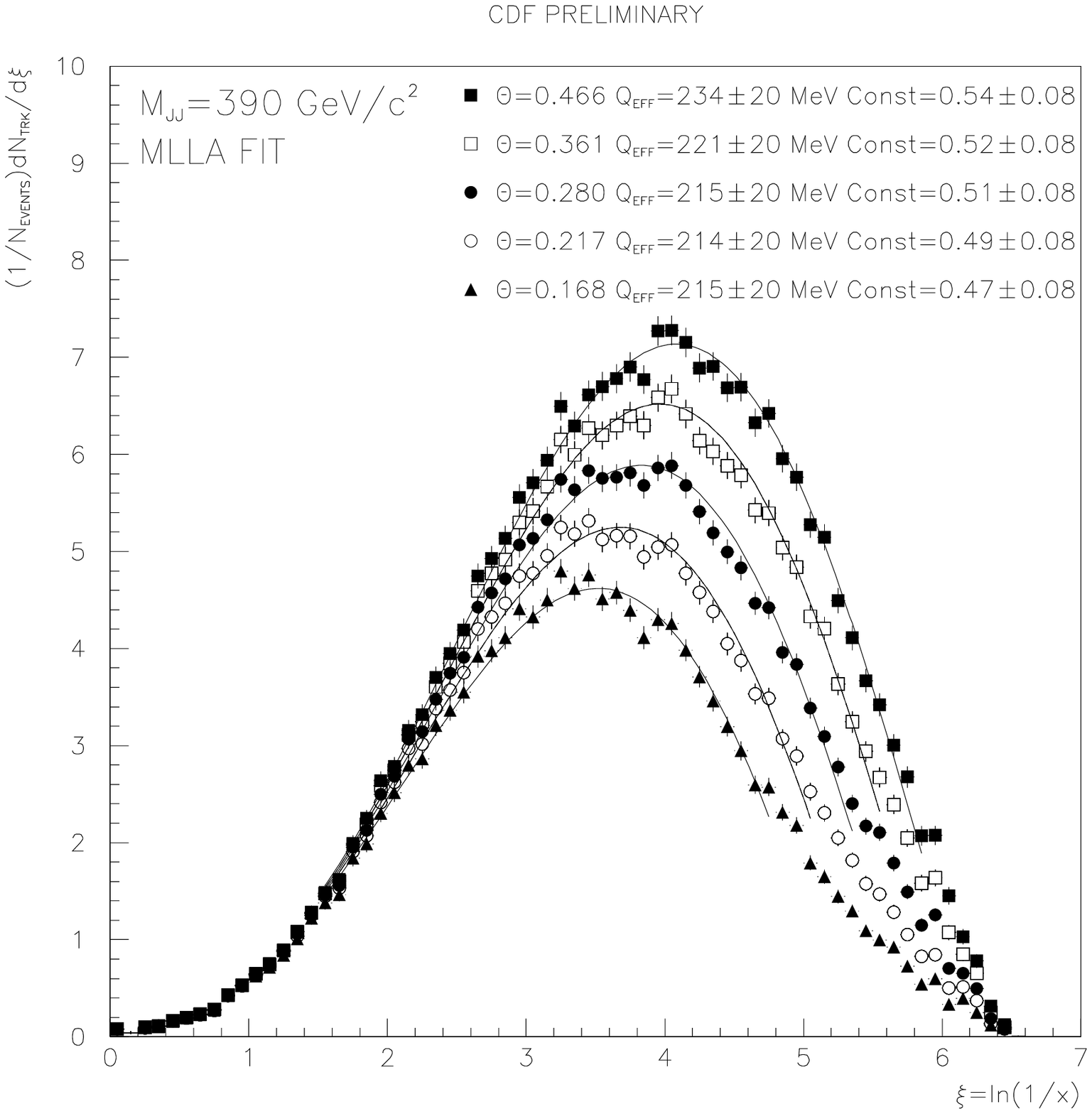,width=0.95\textwidth,height=\textwidth}

\noindent
Fig.1: CDF hump-backed plateau versus an analytic MLLA prediction for
the yield of secondary partons (soft gluons).~\cite{cdf_max}
\end{minipage}
\qquad
\begin{minipage}{0.45\textwidth}
\psfig{figure=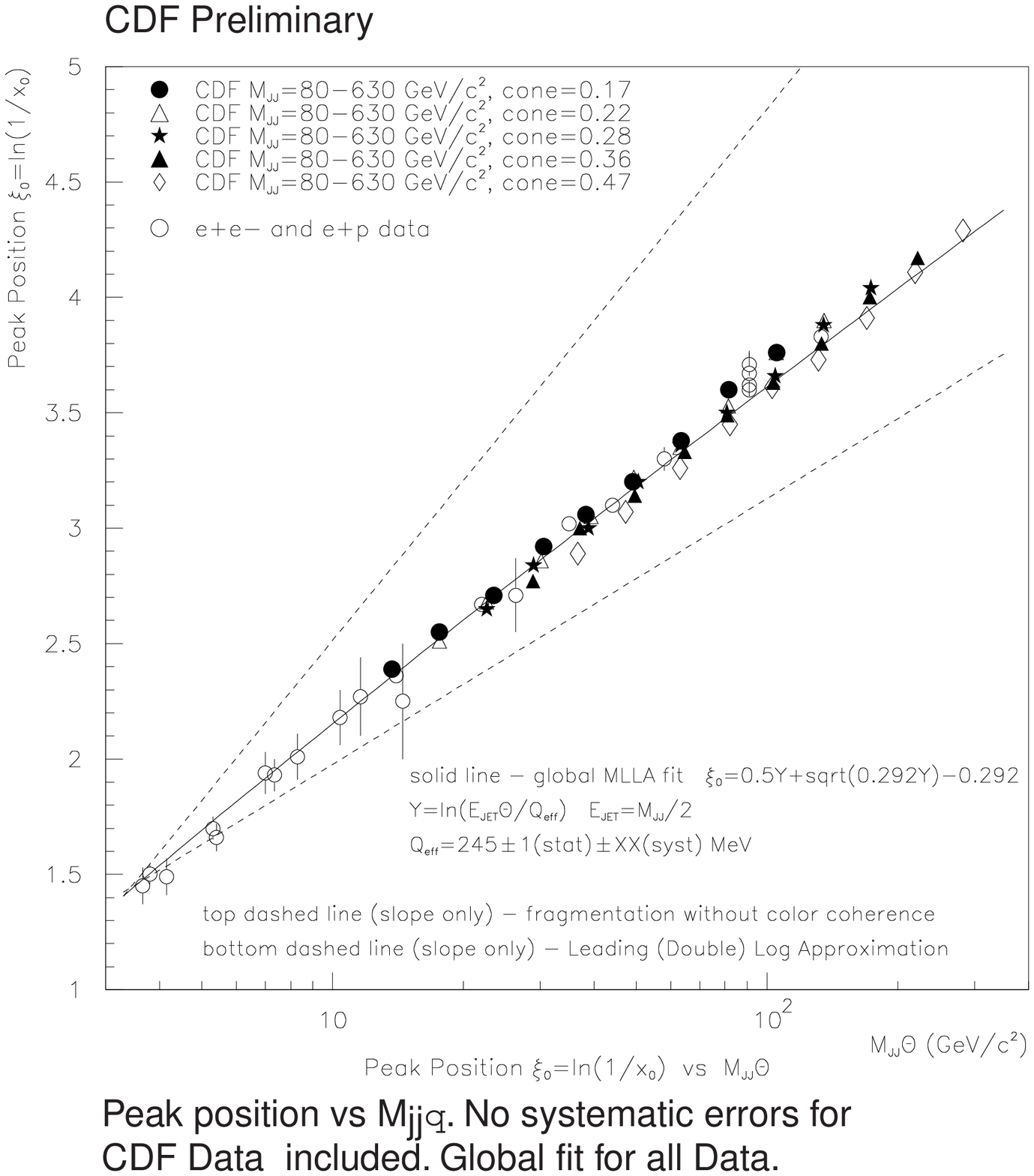,width=0.95\textwidth}

\vspace{2mm}
\noindent
Fig.2: Position of the maximum in the inclusive energy spectra 
versus a parameter-free MLLA prediction.~\cite{cdf_max}
\end{minipage}          
\newpage
                     
The message is, that ``brave gluon counting'', that is applying the
pQCD language all the way down to very small transverse momentum
scales, indeed reproduces the $x$- and $Q$-dependence of the observed
inclusive energy spectra of charged hadrons (pions) in jets.

Even such a tiny (subleading) effect as an envisaged difference in the
position of the maxima in quark- and gluon-initiated humps seems to
have been verified, 15 years later, by the recent DELPHI
analysis.\cite{Bryan_diff,DELPHI_diff}

\subsection{Inter-jet particle flows}
 \wisd{4.5}{"Can you do addition?" the White Queen asked.
   "What's one and one and one and one and one and one and one and one
   and one and one?" "I don't know," said Alice. "I lost count."}
 {\em Through the Looking Glass}

Even more striking is {\em miraculously}\/ successful r\^ole of gluons  
in predicting the pattern of hadron multiplicity flows in the
inter-jet regions --- realm of various {\em string}\//{\em drag}\/
effects.
It isn't strange at all that with {\em gluons}\/ one can get, e.g.,
$1+1\>=\>2$ while  
$1+ 1\>+\> \frac{9}{4} \>=\> \frac{7}{16}$,  
which is a simple {\em radiophysics}\/ of composite antennas, or quantum 
mechanics of conserved colour charges.

\vfill

\noindent
\begin{minipage}{0.45\textwidth}
\psfig{figure=opal_drag.sh,width=1.7\textwidth,height=1.2\textwidth} 

\vspace{0.5cm}

\noindent
Fig.3: Comparison of particle flows in the $q\bar{q}$ valley in
$q\bar{q}\gamma$ and $q\bar{q}g$ 3-jet events versus a parameter-free
analytic prediction based on the soft gluon radiation
pattern.~\cite{OPAL_drag}
\end{minipage}
\qquad
\begin{minipage}{0.5\textwidth}
This particular example of ``quantum arithmetics'' has to do with
comparison of hadron flows in the inter-quark valleys in 
$q\bar{q}\gamma$ and $q\bar{q}g$ (3-jet) events. 
The first equation describes the density of soft gluon radiation
produced by two quarks in a $q\bar{q}\gamma$ event, with 1 standing for 
the colour quark charge. 

Replacing the colour-blind photon by a gluon one gets an additional
emitter with the relative strength 9/4, as shown in the l.h.s.\ 
of the second equation. The resulting soft gluon yield in the
$q\bar{q}$ direction, however, {\em decreases}\/ substantially as a
result of destructive interference between three elements of a
composite colour antenna.

Nothing particularly strange, you might say.  What {\bf is} rather
strange, though, is that this naive perturbative wisdom is being
impressed upon junky 100-200 MeV pions which dominate hadron flows
between jets in the present-day experiments such as the OPAL study
shown in Fig.~3. 
\end{minipage}                                

\vspace{3mm}
\noindent
Another amazing test of this sort was provided by the DELPHI
measurement of the multiplicity of ``(low energetic) tracks emitted
perpendicular to the eveny plane'' in 3-jet
events~\cite{DELPHI_outofplane} which has been found to obediently
follow a simple \PT\ prediction based on coherent soft gluon
radiation.\cite{KLO}
\\[-3mm]
\begin{itemize}
\item The {\em colour field}\/ that an ensemble of hard primary {\bf
    partons} (parton antenna) develops, determines, on the one-to-one
  basis, the structure of final flows of {\bf hadrons}.
\item The Poynting vector of the colour field gets translated into the
  hadron pointing vector without any visible reshuffling of particle
  momenta at the ``hadronisation stage''.
\end{itemize}

When viewed {\em globally}, confinement is about {\em
    renaming}\/ a flying-away quark into a flying-away pion rather
  than about forces {\em pulling}\/ quarks together.

\subsection {Gluons and Gluers}

\underline{Definition:} a {\em Gluer}\/ is a miserable {\em gluon}\/ which 
hasn't got enough time to truly behave like one
because its hadronization time is comparable with its formation time,
$
 t_{\mbox{\scriptsize form.}}\simeq {\omega}/{k_\perp^2} 
\> {\sim}\>  t_{\mbox{\scriptsize hadr.}}\simeq 
\omega R_{\mbox{\scriptsize conf.}}^2
$.
Contrary to respectful \PT\ gluons born with small transverse size,
$k_\perp\gg R_{\mbox{\scriptsize conf.}}^{-1}$, gluers are not
``partons'': they do not participate in perturbative cascading (don't
multiply). 
According to the above definition, gluers have {\em finite}\/
transverse momenta (though may have arbitrarily large {\em
 energies}\/).
Having transverse momenta of the order of inverse 
confinement scale puts gluers on the borderline of applicability of \PT\ 
language, since their interaction strength is potentially large,
$\as(R_{\mbox{\scriptsize conf.}}^{-1})\sim 1$. 
\underline{R\^ole} of gluers is to provide comfortable conditions for {\em
  blanching}\/ colour parton ensembles (jets) produced in hard interactions,
{\em locally}\/ in the configuration space. Gluer formation is a
signal of hadronization process taking place in a given space-time
region. A label to put on the gluer concept might be ---
``{\em A gluer formed $\simeq$ a hadron born}''.   
An \underline{Idea} emerges: To relate {(uncalculable)} 
    Non-Perturbative corrections to {(calculable)} 
    Perturbative cross sections/observables with intensity of {\em
    gluer}\/ emission ({$\as$} in the {infrared} domain).

\section{Power Games 99}

\PT-calculable observables are Collinear-and-InfraRed-Safe (CIS)
observables, those which can be calculated in terms of quarks and
gluons without encountering either collinear (zero-mass quark, gluon)
or soft (gluon) divergences.
Gluers' contributions to such observables are suppressed and are being
rightfully neglected in the pure \PT\ (``logarithmic'') approximation.
These contributions are inversly proportional to
a certain {\em power}\/ of the hardness scale (modulo logs), 
${\delta \sigma^{\NP}}/{\sigma} \propto {\log^q Q}/{Q^{2p}}$.
The corresponding observable-dependent exponents can be inferred from
the analysis of an intrinsic uncertainty in summing up the \PT\ series 
(infra-red renormalons, for an extensive review see~\cite{beneke}). 

Adopting the concept of {\em universality}\/ of \NP\ phenomena one can
{\em predict}\/ the ratio of the magnitudes of power corrections to
different observables belonging to the same $\{p,q\}$ class.

The \PT-approach exploiting gluers allows to go one step further,
namely to relate {\em absolute magnitudes}\/ of genuine \NP\ 
contributions to CIS observables with the intensity of gluer radiation, i.e.\ 
the ``QCD coupling'' at small transverse momentum scales.

\subsection{Phenomenology}
\noindent
\begin{minipage}{0.45\textwidth}

For example, DIS structure functions are expected to
deviate from their perturbative $Q^2$ dependence by terms generally
behaving like $1/Q^2$ {(``twist 4'')}:
$$
F_2(x,Q^2) \simeq F_2^{\PT}(x,Q^2)\left[1+ {D_2(x,Q^2)}/{Q^2}\right].
$$
Comparison of the Power Game prediction~\cite{DW96} with the
data~\cite{VM} allows one to extract the value of the characteristic
\NP\ parameter
$$
 \cA_2 = \frac{C_F}{2\pi}\int_0^\infty dk^2\, \delta\as^{(\NP)}(k)
\>\simeq\> 0.2\>\mbox{GeV}^2\,.
$$ 
\end{minipage}
\qquad
\begin{minipage}{0.45\textwidth}
\epsfig{file=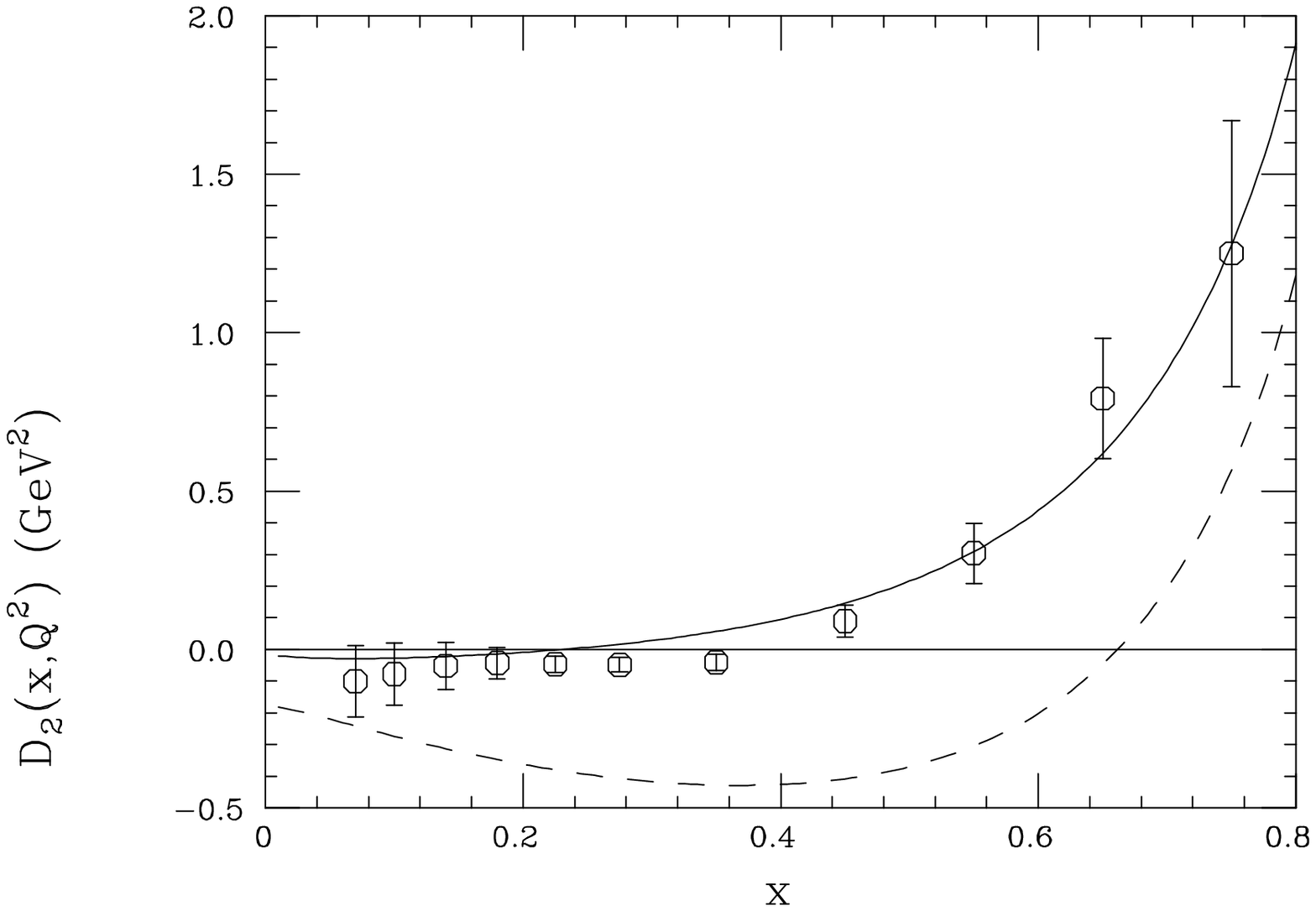,width=7cm}\\[3mm]
Fig.4: $x$-dependence of the $1/Q^2$ contribution to $F_2$ 
\end{minipage}
\vspace{2mm}

\noindent
$\cA_2$ being fixed, 
a parameter-free prediction emerges then for the $1/Q^2$ suppressed
contribution to $F_3$  shown by the dashed curve.

Another example is provided by a variety of observables
including jet shapes (Thrust, $C$-parameter, jet Broadenings,
Oblateness), energy-energy correlation (EEC), $\sigma_L$, etc.\ which
belong to the $p=1/2$ class and thus exhibit numerically large \NP\ 
deviations.

\begin{tabular}{lr}
\epsfig{file=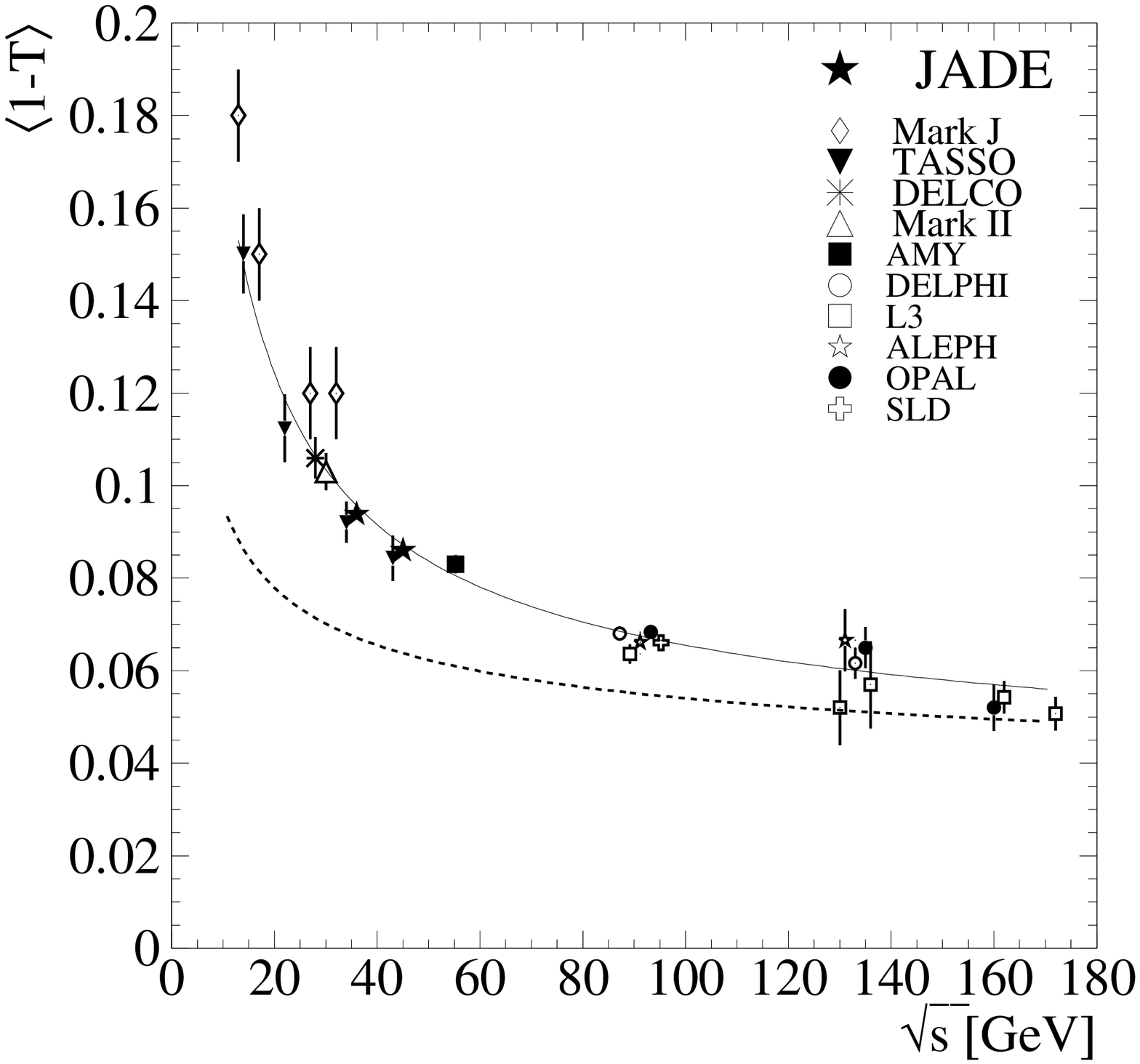,width=0.45\textwidth,height=0.4\textwidth}
&
\epsfig{file=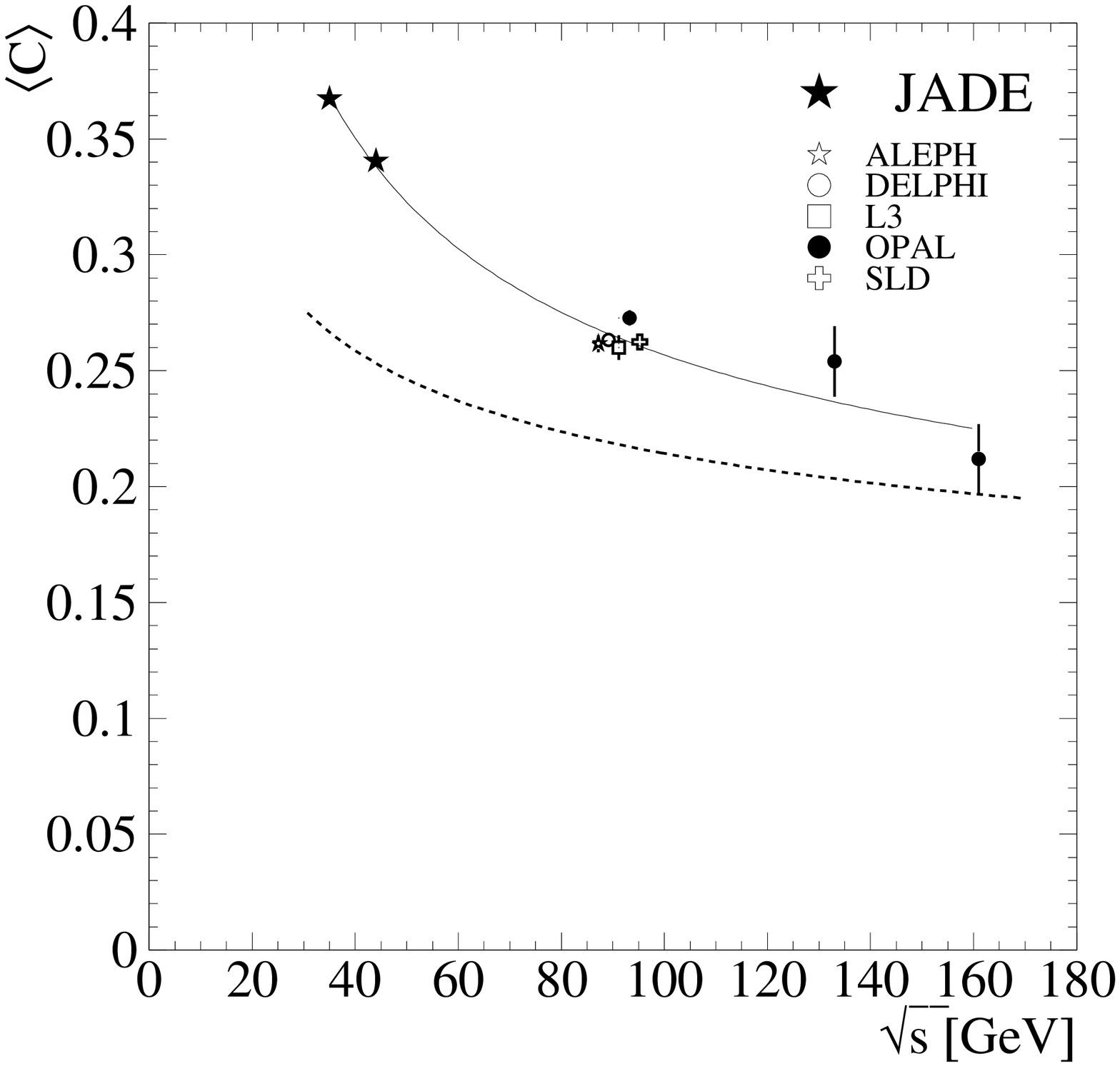,width=0.45\textwidth,height=0.4\textwidth}
\end{tabular}
\begin{center}
Fig.5: Mean Thrust and $C$-parameter in $\epem$ annihilation.\cite{jade}
\end{center}

\noindent
A pure phenomenological study of the
deviation of the mean Thrust and $C$-parameter values from the
corresponding two-loop \PT\ predictions shown by dotted lines in Fig.~5
hints at 
\begin{eqnarray*}
 \delta \lrang{1\!-\!T}^{(\NP)} 
= \lrang{1\!-\!T}-\lrang{1\!-\!T}^{(\PT)}&\simeq& {1\>\mbox{GeV}}/{Q} \>, \\
 \delta \lrang{C}^{(\NP)} = \quad \lrang{C} \>\>-\>\> 
  \lrang{C}^{(\PT)} \quad &\simeq& {4 \> \mbox{GeV}}/{Q}\>,
\end{eqnarray*}
with the power game bet being instead \quad
$
{ \delta \lrang{C}^{(\NP)}}/{ \delta \lrang{1-T}^{(\NP)}} 
\>=\> {3\pi}/{2}
$.

\subsection{Universality of confinement effects  in jet shapes}

\noindent
\begin{minipage}{0.4\textwidth}
  The Power Game grew muscles when it was realised that it can be
  played not only with the $Q$-dependence of the means at stake.  The
  {\em distributions}\/ of shape variables were shown~\cite{shift} to
  be subject to a $1/Q$ {\em shift}, by that very amount that
  describes the genuine \NP\ contribution to the {\em mean}\/ value of
  the corresponding jet shape variable.

  For example, the $C$-parameter distribution (for the values of $C$
  not too close to zero) can be obtained by simply shifting the corresponding
  all-order-resummed purely perturbative spectrum by an amount inverse 
  proportional to $Q$, 
$$
\frac{1}{\sigma}\frac{d\sigma}{dC}(C) \simeq 
\left(\frac{1}{\sigma}\frac{d\sigma}{dC}\right)^{\PT}\!\!\left(C-\frac{D_C}
{Q}\right). 
$$
The corresponding result of a recent JADE analysis is shown in Fig.~6.

The same {\em shift}\/ prescription, and similar high quality
description, hold for other CIS jet observables like Thrust.
\end{minipage}
\quad
\begin{minipage}{0.55\textwidth}
\epsfig{file=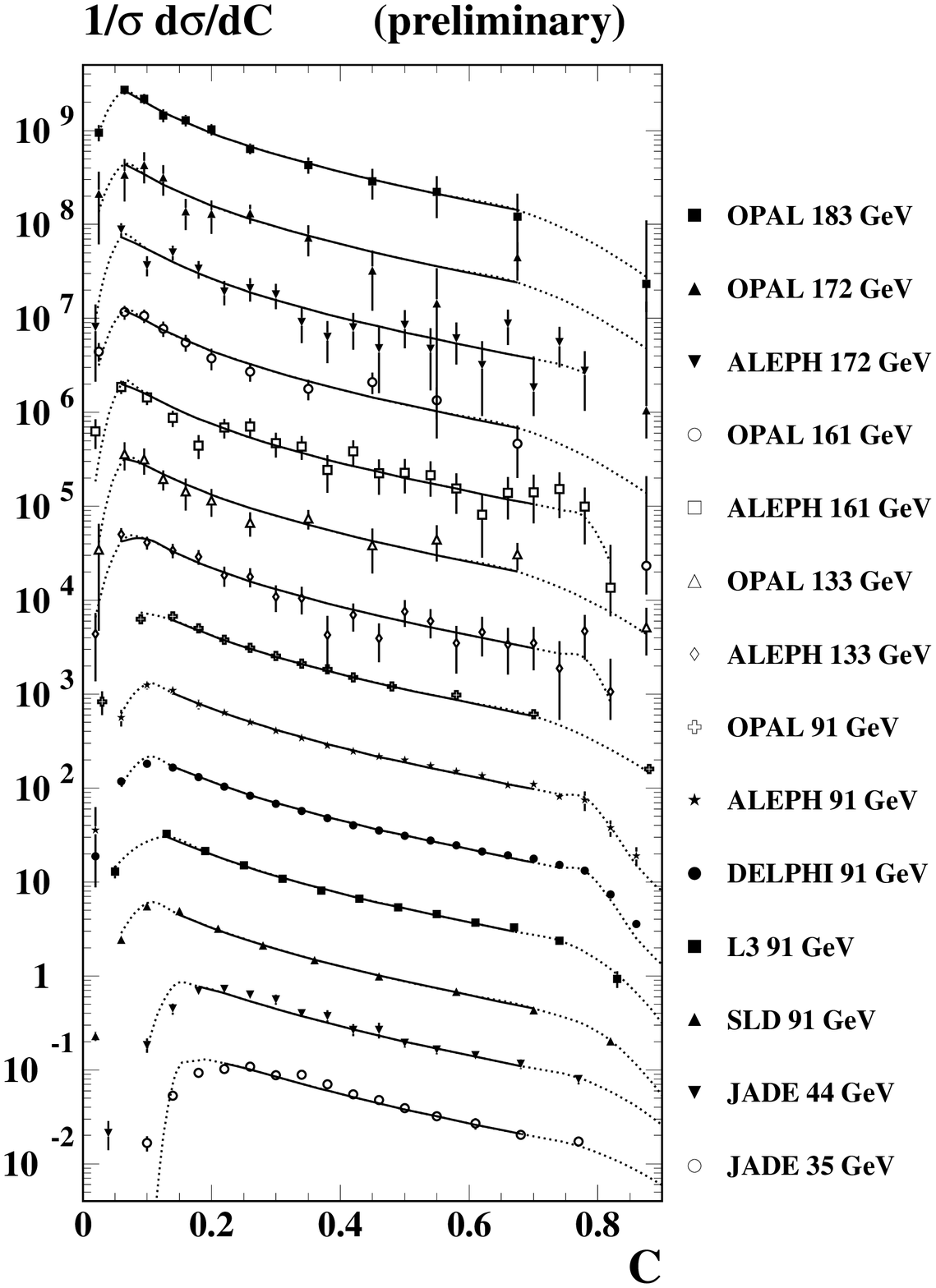,width=\textwidth,height=1.3\textwidth}

\noindent 
Fig.6: $C$-parameter distributions~\cite{tampere}
 versus \PT-spectra {\em shifted}\/ 
by $\delta C\simeq 4\>\mbox{GeV}/Q$
\end{minipage}

\vspace{2mm}

A thrilling story of one important exception was told in
Vancouver.\cite{vancouver} {\em Jet Broadening(s)}\/ defined as a sum
of the moduli of transverse momenta of particles in jet(s) (wrt the
Thrust axis) was first predicted to have a $\log Q$-enhanced \NP\ 
shift, since this \NP\ contribution to $B$ was naturally thought to
accumulate gluers with rapidities up to $\log Q$.

The data however simply could not stand it.\cite{H1,squeeze} 
Fits based on the $\log Q$-enhanced shift were bad and produced too
small a value of $\as(M_Z)$, and the \NP\ parameter $\alpha_0$
inconsistent with that extracted from analyses of the Thrust and
$C$-parameter means and distributions. 

Tragic consequences for the universality belief seemed imminent.

\subsection{Broadening: tragedy, catharsis, lessons}
\wisd{3.5}{{
Raffiniert ist der Herrgott, aber nicht boesartig}}{{
A. Einstein}}

Catharsis came with recognition of the fact that the Broadening
measure ($B$) is more sensitive to quasi-collinear emissions than
other jet shapes, and is therefore strongly affected by an interplay
between \PT\ and \NP\ radiation effects. With account of the
omnipresent \PT\ gluon radiation, the {\em direction of the quark}\/
that forms the jet under consideration can no longer be equated with
the direction of the Thrust axis (employed in the definition of $B$).
As a result of this interplay, the hadron distribution was found to be
not only shifted but also {\em squeezed}\/ with respect to its \PT\ 
counterpart.

\vspace{2mm}
\noindent
\begin{minipage}{0.45\textwidth}
\epsfig{file=btdist-jade35.eps,height=\textwidth,angle=90}
\end{minipage}
\qquad
\begin{minipage}{0.45\textwidth}
\epsfig{file=btdist-opal91.eps,height=\textwidth,angle=90}
\end{minipage}\\[2mm]
Fig.7: Perturbative (dashed) and \NP-shifted/squeezed Total
Broadening distributions.\cite{broad}

Three lessons can be drawn from the Broadening drama.
\begin{itemize}
\item[Pedagogical] lesson the Broadenings taught, was that of the
  importance of keeping an eye on \PT\ gluons when
  discussing effects of \NP\ gluers.  An example of a powerful
  interplay between the two sectors was recently given by the study of
  the energy-energy correlation in $\epem$ in the back-to-back
  kinematics.\cite{EEC} The leading $1/Q$ \NP\ contribution was shown
  to be promoted by \PT\ radiation effects to a much slower falling
  correction, $Q^{-0.32\mbox{-}0.36}$.
 
\item[Physical] output of the proper theoretical treatment was
  restoration of the universality picture: within a reasonable 20\%\ 
  margin, the \NP\ parameters extracted from $T$, $C$ and $B$ means
  and distributions were found to be the same.

\item[Gnostic] output was also encouraging. 
  Phenomenology of \NP\ contributions to jet shapes has shown that it 
  is a robust field with a high discriminative power: it does not
  allow one to be misled by theorists.   
\end{itemize}

\subsection{A step forward: ``shape functions''}
A strong push is being given to the Power Game by the notion of 
Shape Function(s) introduced by G.~Korchemsky and
G.~Sterman.\cite{shapef} 

\noindent
\begin{minipage}{0.45\textwidth}
\epsfig{file=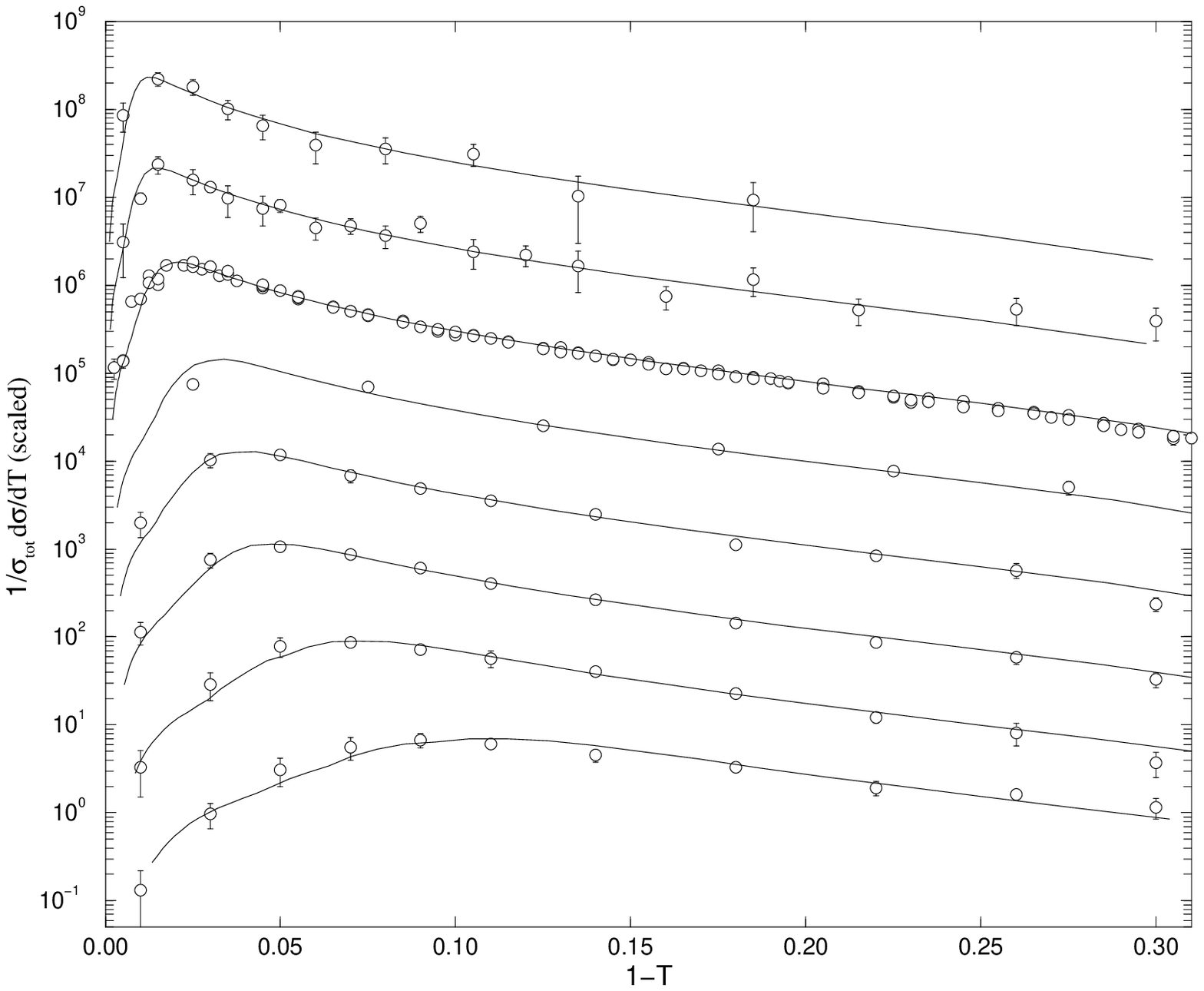,width=1.1\textwidth,height=\textwidth}\\[-5mm]
Fig.8: Thrust distributions at $Q$=14, 22,
  35, 44, 55, 91, 133 and 161 GeV versus Power Game
  prediction~\cite{shapef}
\end{minipage}
\qquad
\begin{minipage}{0.5\textwidth}
  Introducing the {\em distribution}\/ describing the power shift on
  an event by event basis (shape function) makes it possible to lift
  off the condition $1\!-\!T\gg \lrang{1\!-\!T}^{(\NP)}$. A simple
  physically motivated ansatz for such a distribution for the Thrust
  case,
$${
\frac1{\sigma_{tot}}\frac{d\sigma(1\!-\!T)}{dT} = \int_0^{Q(1\!-\!T)}
d\epsilon\, f(\epsilon) \frac{d\sigma\left(1\!-\!T
-\frac{\epsilon}{Q}\right)}{dT}
}$$
with
$$
 f(\epsilon) =  \frac{2\left({\epsilon}/{\Lambda}\right)^{a-1}}
{\Lambda\Gamma\left(\frac{a}{2}\right)}
 \exp\left(-\frac{\epsilon^2}{\Lambda^2}\right) ,
$$
produces a remarkable fit to hadron data shown in Fig.~8.
\end{minipage}

\vspace{2mm}
\noindent
Shape functions for different jet shapes can be related with certain
characteristics of the energy-momentum flow at the hadronization
stage, specific for a given observable.

\subsection{Universality problem}

A detailed discussion of the main problems one faces in establishing
the rules of the Power Game can be found in the proceedings of the 1998
ICHEP.\cite{vancouver} These problems include separation of
power corrections coming from the infrared region from those
determined by the ultraviolet physics, merging (in a renormalon-free
manner) the \PT\ and \NP\ contributions to the full answer, the
problem of splitting the magnitude of the power term into an
observable-dependent \PT-{\em calculable}\/ factor and a {\em
  universal}\/ \NP\ parameter.

The key question is whether the latter is really universal.  The whole
game would have little sense if it were not. Allowing each observable
to have a private fitting parameter we would not learn much about the
way confinement acts in hadronizing ensembles of partons produced in
hard interactions.

Reasonable doubt was expressed in a seminal paper by Nason and
Seymour~\cite{NS} as to whether universality can be expected to hold
for jet shapes which are not truly inclusive observables.  The
configuration of offspring partons in the gluon decay matters for jet
shapes, so that the value of the power term may be affected, in an
observable-dependent way, beyond the leading level in $\as$ (which a
priori is no longer a small parameter since the characteristic
momentum scale is low).

Analyses of two-loop effects in $1/Q$ suppressed contributions have
been carried out for jet shapes in $\epem$ annihilation and DIS. The
output proved to be surprisingly simple.  It was shown that there
exists a definite prescription for defining the so-called ``naive''
one-loop estimate of the magnitude of the power contribution, such
that the two-loop effects of {\em non-inclusiveness}\/ of jet shapes
reduce to a universal, {\em observable-independent}, renormalisation
of the ``naive'' answer by the number known as the ``Milan
factor''.\cite{Milan,DISshapes} This is true for the \NP\ 
contributions in the thrust, invariant jet mass, $C$-parameter and
broadening distributions, for the energy-energy correlation measure,
as well as for other observables subject to linear in $1/Q$
confinement effects.

It is probably the striking simplicity of the resulting prescription
to be blamed for apparently cold reception the ``Milan factor''
enjoyed among theoreticians.  

Verification of the Milan factor prescription is underway.
M.~Dasgupta, L.~Magnea and G.~Smye have undertaken the project of
explicitly calculating the two-loop effects in the \NP\ contribution
to the $C$-parameter distribution.\cite{DaMaSm} The analytical result
they are coming up with has verified the key simplification used in
the original derivation of the Milan factor namely, the soft gluon
approximation. This is good news. The not-so-good news is that the
final expressions for $\cM$ do differ...

\section{Milan factor 2000}
The Power Game as a new theoretical instrument emerged from its
toddler years but has not yet reached respectable teens.  It is
understandable that, being both {\em predictive}\/ and {\em
  verifiable}\/ (the qualities almost extinct nowadays), it attracted
a lot of attention and was developing, in its early days, on a
week-to-week (if not a day-to-day) basis.  
Accelerated childhood tends to be marked by bruises, on the child's
part, and by troubles on the parents'.

A partial history of {\em misconceptions}\/ the advocates of the Power
Game had to muddle through can be found in \cite{vancouver}. Now we
are in a position to enrich this history with a {\em miscalculation}.
An unfortunate omission of a trivial factor in the two-parton phase
space resulted in a wrong value originally derived for $\cM$: the so-called
``non-inclusive'' contribution to the Milan factor, $r^{ni}$, has to
be multiplied  by a factor of 2.
As a result, 
\bea\label{Micorr}
 \cM &=&1+ r_{in}+r_{ni} 
\Longrightarrow 
1+{3.299 C_A}/{\beta_0} 
+ {\mbox{\bf 2}}\times (-0.862 C_A - 0.052n_f)/{\beta_0} \nonumber \\
&=& 1+  (1.575C_A- 0.104n_f)/{\beta_0} 
\>=\> 1.49 \quad \mbox{for}\>\> n_f=3\>, 
\qquad \left(\mbox{instead of}\>\> =1.8\right).
\eea 
The $n_f$-part of the corrected Milan factor Eq.~\ref{Micorr} agrees
with \cite{DaMaSm} and, as the authors point out, also solves the
longstanding discrepancy with the explicit two-loop calculation of the
``Abelian'' ($n_f$-dependent) correction to singlet $\epem$
fragmentation functions ($\sigma_L$) which was carried out by
M.~Beneke, V.~Braun, and L.~Magnea.\cite{BBM}

Refitting jet shape data with the corrected $\cM$ lies ahead.  It will
drive up the \NP\ parameter $\alpha_0$ by about 10\%\ but will change
neither $\as$ nor the present status of the universality pattern.

The situation with universality these days can be viewed as
satisfactory.  It is far from perfect, however.  In particular, there
seems to be a conceptual problem with describing the means and
distributions of those specific jet variables that deal with a certain
single jet rather than the event as a whole. The known cases this
remark applies to, are the {\em Heavy}\/ jet mass and the {\em Wide}\/
jet broadening.  An adequate game strategy for dealing with such ({\em
  less inclusive}\/) observables remains to be found.

A last remark is due concerning the title ``Power Games''. An
ideology and technologies are being developed for describing genuine
confinement effects in various global characteristics of
multi-particle production.  I believe there was a good reason for
calling it a ``game''.  To really enjoy playing one has to follow the
rules (which, by the way, does not contradict the fact that some
entertaining games intrinsically embody bluff).
\\[2mm]
\begin{minipage}{0.55\textwidth}
  In the present context, ``the rules'' means equating ``\PT'' with
  the two-loop prediction and looking upon the rest as being ``\NP''.
  The boundary between \PT\ and \NP\ physics is, to a large extent, a
  matter of convention.  In particular, including an additional loop
  into a ``\PT\ prediction'' (see, e.g.~\cite{AlKa}) or redefining it,
  say, with use of the Borel wisdom,\cite{GG} inevitably affects the
  magnitude of a ``genuine \NP\ contribution''.  Such an elusive
  behaviour of \NP\ effects may appear especially confusing in jet
  shape phenomenology where, according to the Sterman's
  lemma,\cite{sterman_lemma} the NNLO ($\as^3$) effects are perfectly
  capable of mimicking the $1/Q$ behaviour.
\end{minipage}
\quad
\begin{minipage}{0.4\textwidth}
\epsfig{file=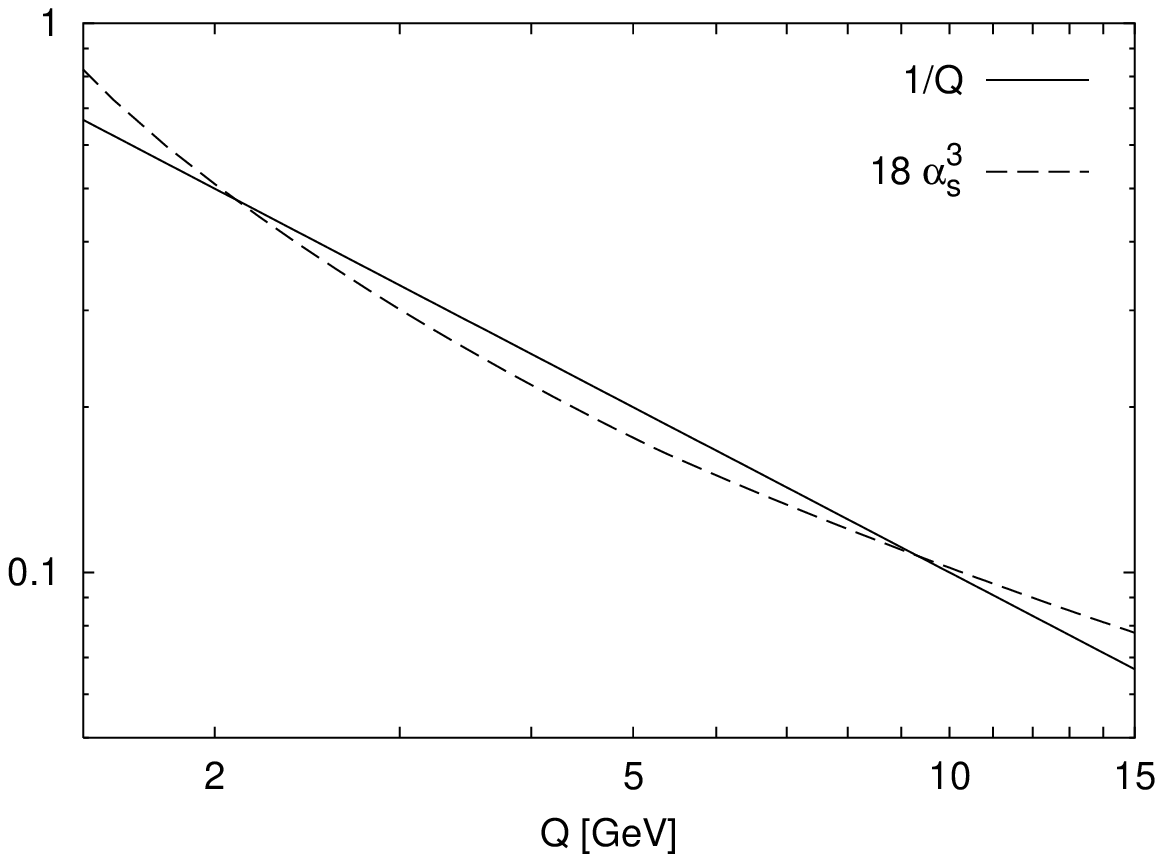,width=\textwidth}

\begin{center}
Illustration of the Sterman's lemma
\end{center}
\end{minipage}

\section*{Acknowledgements}
I am grateful to Mrinal Dasgupta, Pino Marchesini, Gavin
Salam and Bryan Webber for illuminating discussions and help.

\end{document}